# The impact of the pandemic of Covid-19 on child poverty in North Macedonia: Simulation-based estimates

**Marjan Petreski**

University American College Skopje

Finance Think – Economic Research & Policy Institute, Skopje

marjan.petreski@uacs.edu.mk

**Abstract**

The objective of this paper is to estimate the expected effects of the pandemic of Covid-19 for child poverty in North Macedonia. We rely on MK-MOD Tax & Benefit Microsimulation Model for North Macedonia based on the Survey on Income and Living Conditions 2019. The simulation takes into account the development of income, as per the observed developments in the first three quarters of 2020, derived from the Labor Force Survey, which incorporates the raw effect of the pandemic and the government response. In North Macedonia, almost no government measure directly aimed the income of children, however, three key and largest measures addressed household income: the wage subsidy of 14.500 MKD per worker in the hardest hit companies, relaxation of the criteria for obtaining the guaranteed minimum income, and one-off support to vulnerable groups of the population in two occasions. Results suggest that the relative child poverty rate is estimated to increase from 27.8 percent before the pandemic to 32.4 percent during the pandemic. This increase puts additional 19,000 children below the relative poverty threshold. Results further suggest that absolute poverty is likely to reduce primarily because of the automatic stabilizers in the case of social assistance and because of the one-time cash assistance.

**Acknowledgement:** This paper is fully based on the larger study "The Social and Economic Effects of COVID-19 on Children in North Macedonia: Rapid Analysis and Policy Proposals" and its subsequent update, which Finance Think – Economic Research & Policy Institute Skopje and UNICEF Office Skopje produced with a financial support from USAID North Macedonia. The author expresses gratitude to all stakeholders for having agreed to extract the part of the study contained in this paper and to publish it in a journal.

# 1. Introduction

The pandemic caused by the global spread of the coronavirus Covid-19 harmed social, educational and health wellbeing of children, with the most vulnerable being hit the hardest. Children have been impacted directly, through school, extra-curricular and childcare facilities closures, social distancing and confinement, which put a heavy burden on their educational, cognitive and emotional development, with the risk of increasing their anxiety and stress levels. Children have been also impacted indirectly, through the reduction of household incomes, which reduced their material and social wellbeing, impaired access to social and healthcare, while also exposed the hardest hit to risks malnutrition. It is critical to understand that the negative impact of the Covid-19 crisis may be particularly strong for some groups of children including those living in poverty, children with disabilities, children deprived of parental care, children in detention and so on. Furthermore, negative impacts of this scale may extend well beyond the short term spreading childhood poverty across many childhood years or beyond (Elder, 1999).

The experiences of the past economic crises, which hit the children in North Macedonia hard, bear out the expectation that the current crisis will have considerable longer-term unfavorable effects on children. Research and data assessing the Global Financial Crisis of 2009 indicate that children were affected indirectly, as "more than half of the households (56.4 per cent) in North Macedonia were unable to fully provide for the schooling needs of their children, 8.8 per cent of children were experiencing problems with access to education, while 9.3 per cent lacked access to regular health check-ups" (Gerovska Mitev, 2010). Likewise, the poverty of couples with children between 2008 and 2010 increased by 3.4 percentage points, while of other households with children by 5.5 percentage points, amid a smaller rise of headcount poverty of 2.2 percentage points over the same period. Child poverty increased from 34.1 per cent in 2009 to 36.9 per cent in 2010.[1]

The health response to Covid-19 inflicted severe economic and job losses in North Macedonia. With the proclamation of the state of emergency on March 18, 2020 by the President, the technical government[2] introduced strict lockdown measures including a curfew. As a result, part of the economic activity went into induced contraction as bars, restaurants and shopping malls were forced to close. Other part of the economic activity significantly subsided with the closure of borders and restriction of movement, most notably in hospitality and trade. ILO/EBRD (2020) estimated that a full-time equivalent of about 85,000 jobs have been lost (11.5 per cent of total employment) by the mid of the second quarter of 2020. Fortunately, actual job losses have been minimized, particularly due to government's subsidy schemes. However, a significant share of income has been lost.

---

[1] These poverty numbers are based on the Household Budget Survey, which has been abandoned in 2010 with the introduction of the Survey on Income and Living Conditions. Hence, these should not be compared with our later estimates based on SILC.

[2] Appointed on January 3, 2020, with a mandate of 100 days to prepare and conduct fair and democratic elections scheduled for April 12, 2020. This meant that at the time the crisis hit – and the schools closed on March 11, 2020 — North Macedonia did not have a functional parliament, so that the proclamation of the state of emergency by the President was the only legal way to restore the policymaking power of the government.

This very likely exacerbated poverty. In 2019, child poverty (0–17) stood at 27.8 percent, a decline by 1.5 percentage point compared to 2018, yet suggesting that of the total of 448,000 people living under the poverty line about 113,000 were children. The percentage of children who are seriously materially deprived—i.e., child population that cannot afford at least 4 of 9 basic needs—stood at 33.5 percent. Overall, 9.4 per cent of all children in North Macedonia suffered from compound risks (income poverty, material deprivation and living in jobless households) in 2019, nearly four times higher than the EU-28 average. The existing evidence[3] suggests that ethnic Albanian households constitute more than 40 per cent of the poorest quintile, with disposable incomes at only two-thirds of those of Macedonian peers, and the Roma population is not only concentrated in the bottom 40 per cent but is also far below other ethnic groups in labour market outcomes, human capital, and other nonmonetary poverty indicators.

The social and child protection system in North Macedonia went through a comprehensive reform in June 2019 with the objective to further reduce poverty through increase of cash transfers and improved targeting. The law amendments streamlined social and child protection by imposing two major changes: (a) introduction of a guaranteed minimum income for poor households and (b)removal of most of the conditionalities of child allowances, thereby significantly increasing their coverage. Changes included advanced and improved integration of social services, including those aimed at responding to cases of violence against children. Social protection spending has been on an increasing trend reaching 15.5 per cent of GDP in 2019, including the spending on social assistance, but the latter comprised only 1.4 per cent of GDP. Moreover, only quarter of such spending relates to children (Petreski and Petreski, 2018).

The Government deployed two types of policy measures to prevent income loss and maintain the quality of life of citizens: employment-retention measures and additional cash benefits. The key employment-retention measure, first introduced in April-June 2020, was awarding a subsidy of MKD14,500 per worker to the affected companies. It was re-introduced in October-December 2020, with narrower targeting and expanded amounts conditional on the extent of turnover decline. The subsidization of half the employees' social security contributions in the affected companies was abandoned, as it was less attractive to companies compared to the "MKD14,500 per worker" measure (ILO/EBRD, 2020).

The second type of interventions focused on the social transfers income. First, eligibility criteria for guaranteed minimum assistance were temporarily relaxed by (i) introducing means testing based on income in the previous month, rather than the previous three, and by (ii) allowing beneficiaries to own real estate in which they reside, a car older than 5 years and a construction land parcel smaller than 500 m2 – all of which made applicants ineligible before. Finance Think (2020) assessed that the measure had a positive effect on the income of the poorest segments. Second, unemployment benefit eligibility criteria were temporarily relaxed to cover all individuals who lost their job in March and April 2020 regardless of the reasons. Third, the government deployed a significant amount of one-off aid on two occasions: in May 2020 for social transfers income recipients, active registered unemployed, low-pay workers and youth (16–29) in public education; and in December 2020 for passive registered unemployed, low-pension pensioners, youth left out in the first cycle of one off-aid, single parents and some groups of artists. Only the measures introducing one-off cash benefits directly targeted children during the pandemic. With the other social transfers, children have been covered indirectly by targeting workers and households.

---

[3] There is no national statistical disaggregation by ethnicity. Hence, alternative sources are used. The key source about ethnic disaggregation here is the Survey of Quality of Life in Macedonia, which was designed and conducted by Finance Think in May–June 2017. See http://www.financethink.mk/models/survey-on-quality-of-life-in-macedonia-2017/

The objective of this paper is to assess the impact of Covid-19 pandemic on child poverty in North Macedonia. We do the analysis in May 2021, well before the official data on poverty are being published, so as to provide guidance to authorities about potential instant remedies required to prevent longer-term scars. Moreover, we offer a methodological approach based on a tax and benefit microsimulation model, which has seen an increasing utilization with the outbreak of the pandemic. Hence, these two constitute contributions to the current sparse of knowledge about how the pandemic affects child poverty.

The reminder of the paper is organized as follows. Section 2 presents the analytical framework, namely the details of the simulation exercise, assumptions and data used. Section 3 presents the results and offers a discussion. Section 4 concludes and offers a policy-relevant discussion.

## 2. Analytical framework

This assessment of the child poverty under Covid-19 uses the MK-MOD Tax and Benefit Microsimulation Model for North Macedonia. It is a static model where individual behavior (labor-market activity, employment, childcare, saving, etc.) is assumed to be exogenous to the tax-benefit system. It belongs to the family of "standard" static models where individuals/households choose to supply labor (hours of work) until the point where the "marginal disutility of work equals the marginal utility of disposable (net-of-tax) income." (Saez, 2010, p.180). In this setting, taxes and social transfers affect the labor-market behavior by changing the relative value of work vs. leisure.

MK-MOD allows for simulation of social assistance, child allowances, unemployment benefits, direct taxes and social security contributions. For the purpose of the modelling exercise in this study, we rely on a limited set of features of MK-MOD, the primary reason being that we reset the model on SILC 2019[4]. Hence, we simulate wages from employment, including informal wages; income from self-employment; and their associated tax and social contributions wedges; then, guaranteed minimum income and child and educational allowances. The following categories of income and associated taxes are not modelled but taken as reported in 2019 SILC: income from capital (rents and dividends); pensions; unemployment benefits; and inter-household cash transfers.

---

[4] The original MK-MOD was built on the basis of Finance Think's own Quality of Life Survey 2017.

There are, however, two important caveats. The first one is related to the unemployment benefit, which we do not simulate because it assumes information about past tenure and salary of the employee who lost a job. Moreover, unemployment benefit eligibility is quite stringent and if termination of the contract occurs at the request of the employee or by mutual agreement, the unemployed person loses eligibility. None of this information is contained in SILC, which prevents simulation. The second caveat relates to the reformed social assistance scheme, as introduced in Section 1; the reform enacted in May 2019 is captured in SILC 2019. However, since the reform was implemented half way through 2019, and given that it is reasonable to assume that there was a time-lag between implementation and its effects on the beneficiaries, the simulated changes combine the effects of the 2019 reform and of the relaxation of the cash benefits eligibility criteria in response to COVID-19.

In order to simulate the effect of COVID-19 on household incomes, we use parameters of how incomes behaved during the pandemic from another survey, namely, the Labor Force Survey (LFS), which is conducted quarterly in North Macedonia. At the time of writing of this update, LFS Q1-Q3 was published. LFS has two key income sources: from wages and from self-employment. Therefore, we make use of the observed changes over Q1-Q3 of 2020, compared to 2019, to capture the actual effect of the crisis (which combines the shock and the government response, primarily through the key measure of wage subsidies).

We use this information in the following way. For wages, we define a cell by the branch at two-digit NACE Rev.2 (a total of 89 branches), sex (males and females) and age group (youth 15-24, adults 25-49, elderly adults 50-64), which leads to 534 cells. Then, for each cell – e.g., employees in computer programming who are young and male – we calculate the total income from wages in the observed periods: 2020 (first three quarters, scaled to reflect the year) and 2019 (entire year). For each cell, we obtain the change in wage income between the two periods. For the cells for which the number of employees in 2019 has been less than 1,000, we arbitrarily assign a no-change number of wage income, because such small cells face large standard errors, and the actual numbers may be driven by particular observations. For self-employment income, we calculate the changes in income for 21 sectors at one-digit NACE Rev.2. Hence, we apply the same approach, but our cells here are aggregately defined, since splitting the sample of self-employed further than this leads to large standard errors. Therefore, we have at our disposal the observed changes during the pandemic in wage income in 534 cells and observed changes in self-employment income in 21 cells. We use them to simulate how the income observed in 2019 SILC potentially behaved during the pandemic of 2020.

We check the general validity of our procedure by comparing the behavior of wage and self-employment income from LFS and after that we conduct the simulation in SILC. The results are presented in

**Table 1**. In general, the simulated changes are of a comparable magnitude with the observed changes, which proves satisfactory robustness of the approach.

**Table 1 – Simulated versus observed changes in wage and self-employment income during COVID-19 pandemic**

|  | SILC (simulated) | LFS (observed) |
|---|---|---|
| **Wage income** | 5% | 9.8% |
| **Self-employment income** | -11.6% | -10.7% |

*Source: Authors' calculations based on SILC 2019, LFS 2019 and 2020, and MK-MOD.*

The simulation of the guaranteed minimum assistance (GMA) is based on the assumption that the removal of the income from rent component as criterion mimics the relaxation of the eligibility criteria. On the obtained number of GMA recipients, we simulate the extension of the energy supplement from 6 to 12 months. More importantly, the losses in wage/self-employment income increase the number of households eligible for GMA (the so-called, automatic stabilizers); the means testing was done on income in the last month, instead of the last three, but over time this lost general importance given that the crisis has already lasted for 10 months by the time of writing.

Finally, we simulate one-time cash assistances that the government pursued on two occasions:

- In May 2020, assistance of MKD9,000 to persons older than 18 living in households receiving social assistance (GMA and other allowances) and to active registered job seekers; one-time cash assistance of MKD3,000 to employees whose income originates solely from employment and which did not exceed MKD15,000 monthly; and one-time cash assistance of MKD3,000 to regular public education pupils and students aged 16–29.
- In December 2020, the one-off support was aimed at the passive jobseekers with low income (not exceeding MKD15,000 monthly in 2020); pensioners with pension income lower than MKD15,000 monthly; and some other specific categories (like film artists, singers, etc.).

## 3. Results and discussion

Based on the methodology and assumptions outlined above, a new set of household income data that we call "post-COVID-19 scenario" was produced. Based on these data, the post-COVID-19 child poverty rates were calculated taking into account the differential effects of wage income, self-employment income, social transfers income, and one-off payments. We have also disaggregated child poverty rates by some individual and household characteristics.

**Table 2** presents the effects of COVID-19 on child poverty by contrasting 2019 headline child poverty rates with the post-COVID-19 estimates. Three child poverty indicators are used: the relative one based on the share of the child population living in households whose income has fallen below the 60th percentile of the median equivalent income; and two absolute poverty rates – extreme-low and upper-middle-income poverty thresholds. For the discussion of caveats related to the poverty lines used see Petreski et al. (2020a). Columns (2)–(5) refer to the impact of each factor on the pre-COVID-19 poverty rate, considered separately, while column (6) presents the simulation of the impact of all four factors taken together.

COVID-19 is projected to strongly affect child poverty. The relative poverty rate is estimated to have increased from 27.8 percent before the pandemic to 32.4 percent during the pandemic. This increase of 4.6 percentage points would put additional nearly 19,000 children in North Macedonia below the relative poverty threshold.[5] Results further suggest that absolute poverty is likely to reduce primarily because of the automatic stabilizers in the case of social assistance and because of the one-time cash assistance.

**Table 2 – COVID-19's effect on child poverty**

|  | Pre-COVID-19 | Post-COVID-19 | | | | Assessment for 2020 |
|---|---|---|---|---|---|---|
|  |  | Individual impact of considered factors | | | | |
|  | Actual 2019 | Impact of wage income decline (incl. informal wages) | Impact of self-employment income decline | Impact of social assistance relaxation | Impact of one-time cash assistance |  |
|  | (1) | (2) | (3) | (4) | (5) | **(6)** |
| **Relative poverty (below 60% of the equiv. median income)** | 27.8% | 30.9% | 31.3% | 28.4% | 28.3% | 32.4% |
| **Absolute poverty, below extreme-low-income threshold** | 1.5% | 2.1% | 2.8% | 0.5% | 0.4% | 0.6% |
| **Absolute poverty, below upper-middle-income threshold** | 8.1% | 8.5% | 9.3% | 8.1% | 7.1% | 7.5% |

*Source: Authors' calculations based on SILC 2019 and MK-MOD.*

Table 2 presents further interesting insights. The incremental effect of the declining wage income (including the severe reduction of wages from informal jobs, column 2) is assessed at about 3.1 percentage points, while the effect of the decline of self-employment income (column 3) – at 3.5 percentage points. This suggests that self-employed individuals are more frequently hovering around the poverty threshold, and a large-scale shock, as is the one from Covid-19, easily throws them into poverty. This is also not surprising given that a large share of the self-employed are found in the hardest hit sectors, like retail trade, personal services (hairdressers and cosmetics salons), bars, restaurants and transport.

---

[5] According to SSO population estimates, the number of children 0–17 in 2019 in North Macedonia was 407,865.

We assess that the relaxed GMA criteria in response to Covid-19 reduces absolute poverty from 1.5 to 0.5 per cent, assuming all eligible poorest of the poor segments obtain the assistance. This result is consistent with the fact that the majority of incomes below the upper middle-income threshold are social incomes. However, the reasons for the failure to eradicate extreme poverty should be again sought in the property ownership eligibility criteria, as well as in the likely inadequacy of assistance amounts for very large families.

Column (5) presents the incremental impact of the one-time cash support which, interestingly, has a significant effect on absolute poverty. However, one needs to be cautious in the interpretation, because for most part, these cash disbursements top up already low incomes – either from social assistance or from low-pay employment – and hence cumulatively show up powerful in eradicating extreme poverty. They are, therefore, properly targeted, but alone they likely cannot eradicate poverty.

Overall, income losses due to Covid-19 are found to primarily exacerbate relative child poverty, though to some extent also the absolute poverty as per the upper-middle-income threshold. Relaxed GMA criteria and one-off cash disbursements are softening the impact on child poverty mainly in the left tail of the income distribution, hence affecting primarily absolute poverty rates, which is determined by the income threshold for the GMA eligibility and the targeting of one-time cash support.

These estimates assume that the effect of the COVID-19 crisis in the fourth quarter of 2020 (unobserved at the time of writing of this paper) followed the pattern of the first three quarters. This may be a reasonable assumption, given that the "MKD14.500 per worker" measure was repeated in October 2020. Even though it had a narrower targeting, it nonetheless helped to sustain incomes and jobs. Self-employment income may have declined similarly to the trend in the previous quarters, given some types of small businesses – such as wedding restaurants, playrooms and language schools – remained closed, while others – such as restaurants and cafes – continued to face restrained demand. The government's fourth package of response measures also included small grants for the small businesses that remained shut, but it is more likely that they would be used for covering the accrued fixed costs (most notably rents) than for self-employment income. Given these uncertainties related to income developments in Q4-2020, we introduce a confidence interval of 20 percent on both sides and re-calculate the total relative poverty.

**Figure 1** presents the results and suggests that regardless of whether the impact of COVID-19 on income will lessen or increase, the estimated number of additional children facing relative poverty is expected to rise significantly during the pandemic. Namely, if the crisis impact in Q4-2020 is 20 percent lower than the overall impact in 2020, then the child poverty rate would have subsided to 31.9 percent; and if it was 20 percent higher than the overall impact, then the child poverty rate would have increased to 33.2 percent. With this interval, the number of children thrown into relative poverty due to the COVID-19 crisis lies between 17,000 and 22,000.

**Figure 1 – Uncertainty around the relative child poverty rate estimate (2020)**

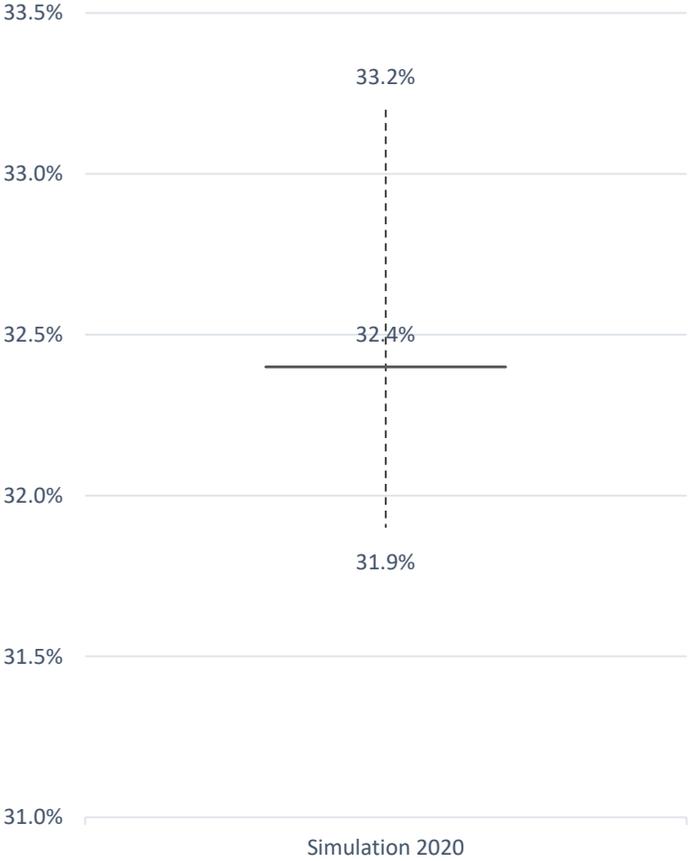

*Source: Authors' calculations based on SILC 2019 and MK-MOD.*

We proceed by disaggregating the COVID-19's effects on child poverty by individual and household characteristics.

**Figure 2** presents child poverty rates by gender and age of the child. Girls fall into poverty more frequently than boys, but in terms of relative poverty, the pandemic affects boys disproportionally stronger than girls. Conversely, boys experience more favorable effects of expanded social transfers on extreme poverty. In 2020 we found fewer differences between the sexes. In terms of age, COVID-19's effect is spread among all the sub-categories. The relative poverty surges by about 4 percentage points, while the absolute poverty rate at the upper-middle-income threshold declines for the 0–5 and 6–14 sub-groups.

**Figure 2 – COVID-19's effect on child poverty, by gender and age**

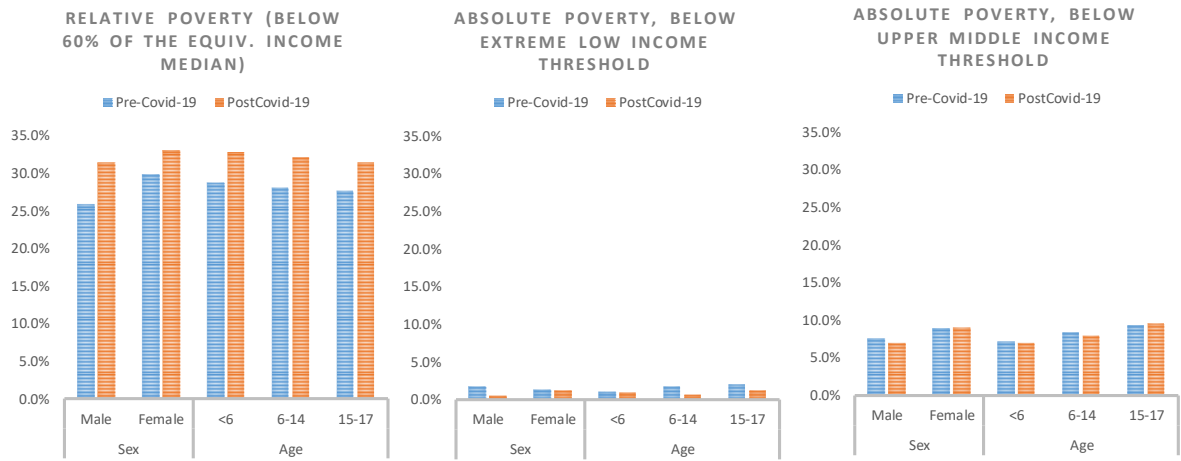

*Source: Authors' calculations based on SILC 2019 and MK-MOD.*

**Figure 3** looks at two household characteristics: whether the household has 3 or more children, and the average education of the adult members. The relative poverty of children in multi-child households is astonishingly high, and COVID-19 aggravates it. However, the poverty increasing effect of the pandemic is not tilted towards households with 3+ children; quite the opposite, extreme poverty is more intensively eradicated among such households. Child poverty is significantly more prevalent in households where the average education is primary or less, and COVID-19 further intensifies it. On the other side of the spectrum, when adult members of the household have an average of tertiary or higher education, child poverty grows post-COVID-19 but remains very low. However, the relaxation of social assistance eligibility criteria and the one-off aid have likely properly targeted those with high intensity of poverty, i.e., households with 3+ children and with low average education, as it is likely that they were at the same time the households with no or low work intensity, which constituted eligibility for such assistance.

**Figure 3 – COVID-19's effect on child poverty, by household characteristics**

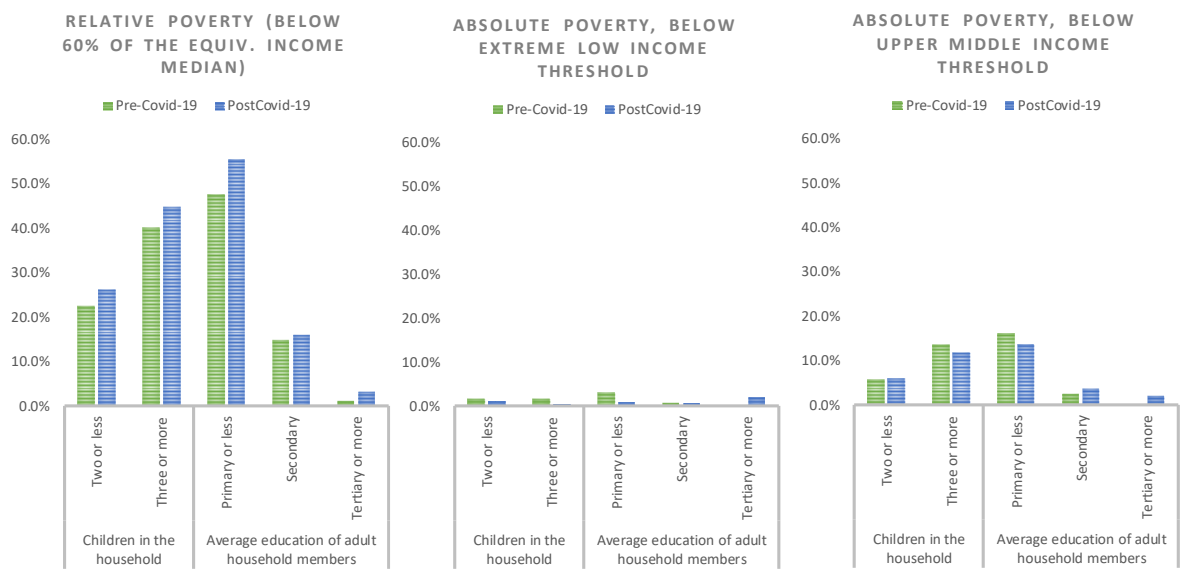

*Source: Authors' calculations based on SILC 2019 and MK-MOD.*

# 4. Conclusion and policy recommendations

The objective of this paper was to estimate the expected effects of the pandemic of Covid-19 for child poverty in North Macedonia. We rely on MK-MOD Tax & Benefit Microsimulation Model for North Macedonia based on the Survey on Income and Living Conditions 2019. The simulation takes into account the development of income, as per the observed developments in the first three quarters of 2020, derived from the Labor Force Survey, which was already available at the time of writing of this paper. Based on such observed data, we follow the pattern of income, which incorporates the raw effect of the pandemic as well the government response. In North Macedonia, almost no government measure directly aimed the income of children, however, three key and largest measures addressed household income: the wage subsidy of 14.500 MKD per worker in the hardest hit companies, relaxation of the criteria for obtaining the guaranteed minimum income, and one-off support to vulnerable groups of the population in two occasions.

The results suggest that the relative child poverty rate is estimated to increase from 27.8 percent before the pandemic to 32.4 percent during the pandemic. This increase of 4.6 percentage points would put additional nearly 19,000 children in North Macedonia below the relative poverty threshold. Results further suggest that absolute poverty is likely to reduce primarily because of the automatic stabilizers in the case of social assistance and because of the one-time cash assistance.

For better tailoring of the system for crises, particularly of such unprecedented nature, the government may consider designing a temporary basic income (TBI) scheme. The objective of a TBI is to top up incomes of people with livelihoods below a vulnerability-to-poverty threshold, which is usually above the value of the poverty line. Technical details may be seen in Petreski et al. (2020b). For example, one of the simulated options of TBI awards is a cash transfer equivalent to a fourth of the median household per capita income with an estimated monthly cost of about EUR40 million, about 2.5 times the current spending on social assistance. Designed this way, TBI spending is strongly pro-children, i.e., assigns a larger share of the funds to households with children (53.4 per cent) than their share in the population (39.3%). For the households with three and more children, the effect is even stronger: 13.2 per cent share of TBI compared to 6.9 per cent participation in population. As a result of such a measure, child poverty would be expected to decline by as much as one third.

Over the long term, the government may consider the universal child guarantee. This will involve removing the current income and property eligibility conditions for child allowance, and increasing the child allowance for GMA recipients to the level where the sum of the GMA (including the energy subsidy), the education allowance and the child allowance reaches the minimum wage. Such an endeavor needs to be: i) combined with interventions within the tax system in order to mobilize funds for its financing; and ii) carefully weighed against the possible labor-market distortions it may entail. The authorities need to carefully examine and take on board the experience of the failed progressive income tax reform of 2019. It is key to, first, work on enlarging the tax base and improving collection efforts, so as to collect more revenues with the same tax rates. Second, reforms of the profit tax and transfer prices, as well as the revision of specific tax exemptions, may provide further strengthening of the tax base. Finally, an income tax reform introducing progressivity for incomes from labor and capital, as well as taxing currently not taxed capital gains and interest, would be a logical next step.

Given that more than a quarter of the citizens who meet the relaxed GMA eligibility criteria have not utilized this cash benefit, it is vital to design outreach programs that would be put in place in emergency situations to inform all potential beneficiaries, including the households left furthest behind, of the changes in the provided social services and cash benefits and their delivery in times of crises. Even though the December 2020 amendments to the Social Protection law have introduced certain elements of shock responsiveness into the system, there is still a need to develop standard operating procedures and contingency plans to ensure rapid mobilization of human and other resources, as well as professional development and support programs enabling the social workforce to continue and even scale up work during emergency situations. Additionally, setting spending targets in the three child-related sectors—education, healthcare and social protection— for the next years may help in the medium-term planning and in the mitigation of the effect of COVID-19 crisis on multidimensional child poverty. There will be higher needs for additional resources (human resources, materials, soft skills, digitalization) for coping with any lingering effects of COVID-19 in early childhood education, schooling and healthcare for children.

# 5. References

Elder Jr., G.H. (1999) *Children of the Great Depression: Social Change in Life Experience*. Westview Press.

Finance Think (2020) To what extent will COVID-19 increase poverty in North Macedonia? Policy Brief 43.

Gerovska Mitev, M. (2010) The Well-being of Children and Young People in Difficult Economic Times. Skopje: UNICEF.

ILO/EBRD (2020) Covid-19 and the World of Work. North Macedonia - Rapid Assessment of the Employment Impacts and Policy Responses. Geneva: International Labor Organization.

Petreski, B. and Petreski, M. (2018) Analysis of the Public Spending on Education and on Social Protection of Children in the Country. Finance Think Policy Study No. 20.

Petreski, M., Petreski, B., Dimkovski, V., Stojkov, A., Tomovska-Misoska, A. & Parnardzieva-Zmejkova, M. (2020b) Bridging measures to alleviate COVID-19 consequences: Design proposal, cost and key effects. Finance Think Policy Studies 2020-12/32, Finance Think - Economic Research and Policy Institute.

Petreski, M., Petreski, B., Tomovska-Misoska, A., Parnardzieva-Zmejkova, M., Dimkovski, V., Gerovska-Mitev, M. & Morgan, N. (2020a) The Social and Economic Effects of COVID-19 on Children in North Macedonia: Rapid Analysis and Policy Proposals. Finance Think Policy Studies 2020-07/30, Finance Think - Economic Research and Policy Institute.

Saez, E. (2010) Do Taxpayers Bunch at Kink Points? *American Economic Journal: Economic Policy*, 2(3): 180–212.